\documentclass[reprint,superscriptaddress,amsmath,amssymb,aps,pra,floatfix]{revtex4-1}

\usepackage{graphicx}
\usepackage{dcolumn}
\usepackage{bm}
\usepackage{hyperref}
\usepackage{natbib}
\usepackage{xcolor}
\usepackage{siunitx}

\begin{document}

\title{Synthesis of Cobalt Grown from Co-S Eutectic in High Magnetic Fields}

\author{Steven Flynn}
\affiliation{Department of Physics, University of Florida, Gainesville, FL 32611-8440, USA}
\author{Caeli L. Benyacko}
\affiliation{Department of Physics, University of Florida, Gainesville, FL 32611-8440, USA}
\author{Mat\'u\v{s} Mihalik}
\email[Email: ]{matmihalik@saske.sk}
\affiliation{Institute of Experimental Physics, Slovak Academy of Sciences, 
Watsonova 47, Ko\v{s}ie 040 01, Slovakia}
\author{Jared Lee}
\affiliation{Department of Physics, University of Florida, Gainesville, FL 32611-8440, USA}
\author{Fuyan Ma}
\affiliation{Department of Chemistry, University of Florida, Gainesville, FL 32611-7200, USA}
\author{Michael E. Bates}
\affiliation{Department of Physics, University of Florida, Gainesville, FL 32611-8440, USA}
\author{Shubham Sinha}
\affiliation{Department of Physics, University of Florida, Gainesville, FL 32611-8440, USA}
\author{Khalil A. Abboud}
\affiliation{Department of Chemistry, University of Florida, Gainesville, FL 32611-7200, USA}
\author{Marian Mihalik}
\affiliation{Institute of Experimental Physics, Slovak Academy of Sciences, 
Watsonova 47, Ko\v{s}ie 040 01, Slovakia}
\author{Mark W. Meisel}
\email[Email: ]{meisel@ufl.edu}
\affiliation{Department of Physics, University of Florida, Gainesville, FL 32611-8440, USA}
\affiliation{National High Magnetic Field Laboratory, University of Florida, Gainesville, FL 32611-8440, USA}
\affiliation{Institute of Physics, Faculty of Science, 
P.~J.~\v{S}af\'{a}rik University, Park Angelinum 9, 040~01 Ko\v{s}ice, Slovakia}
\author{James J. Hamlin}
\email[Email: ]{jhamlin@ufl.edu}
\affiliation{Department of Physics, University of Florida, Gainesville, FL 32611-8440, USA}

\date{\today}

\begin{abstract}
Samples of Co were grown directly in the ferromagnetic state under equilibrium 
conditions using a cobalt sulfide flux. 
Magnetic fields up to 9 T were applied during growth, and isolated Co 
products exhibit progressively elongated morphologies, 
from cubes to rectangular rods to needle-like tendrils with poorly-defined facets. 
The degree of elongation of the major axis was found to 
correlate with magnetic field direction, strength, and gradient.  
Two-dimensional X-ray diffraction data indicate some 
level of polycrystalline-like samples, 
and quantitative analyses (Le Bail and Rietveld) of the one-dimensional data 
confirm the presence of hcp and fcc phases. 
The magnetic responses indicate a partial alignment of the magnetic easy-axis of the 
hcp phase along the magnetic field present during growth.   
\end{abstract}

\maketitle

\section{Introduction}
The application of high magnetic fields during materials synthesis and processing 
at high temperatures has been the subject of research for at least four decades 
\cite{Steiner1989,Rango1991,Beaugnon1993,Koch2000,Yamaguchi2007,Sun2012,Cao2020,Wei2021}. 
An initial impression is the presence of the magnetic field is not expected to 
have a significant effect because the ratio of the magnetic to thermal energy 
is estimated to be small, 
\emph{e.g.},~$\mu_{\mathrm{eff}} B / k_{\mathrm{B}} T \lesssim 0.1$ 
for $\mu_{\mathrm{eff}} = 5\,\mu_{\mathrm{B}}$, $B = \SI{10}{\tesla}$, 
and $T = \SI{300}{\kelvin}$, where $k_{\mathrm{B}}$ is the Boltzmann constant and 
$\mu_{\mathrm{B}}$ is the Bohr magneton.  Contrarily, an increasing 
number of materials have been reported to possess altered properties 
when being synthesized or processed at extreme magnetothermal conditions 
\cite{Jaramillo2005,Koyama2011,Laughlin2019,Weiss2021,Wang2022}.

\begin{figure}[b!]
\centering
\includegraphics[width=3.375in]{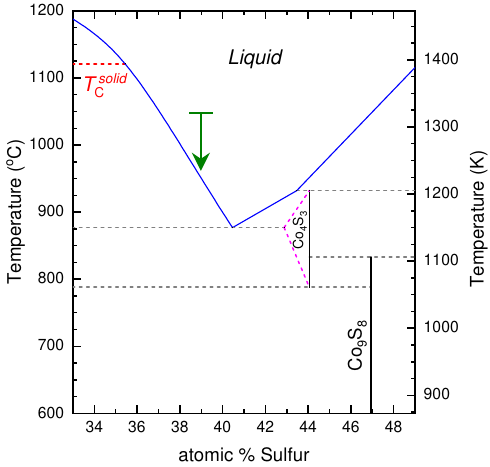}
\caption{The Co-S binary alloy phase diagram, adapted from \cite{ASM1990} which   
is reproduced and given in \cite{ASM2016}, presumably based on aspects of another 
compilation \cite{Rau1976}, is shown in the region of interest 
for this work.\footnote{It is important to note 
this phase diagram was generated in the absence of 
an applied magnetic field.}  The liquidus boundary is shown in blue, and the Curie temperature 
for pure solid cobalt is shown in red. The expected path of our 
composition Co$_{61}$S$_{39}$ is shown by the green arrow, whose base is located 
at the hold temperature of \SI{1050}{\celsius} and body spans the slow-cooling 
path to nominally \SI{950}{\celsius}, where crystallization occurs as the 
liquidus line is encountered.}
\label{fig:1}
\end{figure}

The present work addresses the specific endeavor to grow and characterize 
single crystalline Co grown at temperatures below the Curie temperature 
of the solid by using a cobalt-sulfide flux, 
as suggested by Canfield and coworkers in 2012 \cite{Lin2012,Lin2013,Canfield-2020}, Fig.~\ref{fig:1}.
Possessing the highest known Curie temperature  
($T_{\mathrm{C}} = \SI{1121}{\celsius} = \SI{1394}{\kelvin}$) of any single element, 
cobalt is an ideal model whose structural and magnetic properties are well-studied 
\cite{Honda1926,Kaya1928,Nishiyama1934,Myers1951,Sucksmith1954,Rau1976,ASM1990,ASM2016}.
In fact dating back to the early 2000s, Beaugnon and collaborators have generated 
Co products from \hbox{Co-B}~\cite{CoB-Gaucherand2004,CoB-Wang2015,CoB-He2018,CoB-He2019a,
CoB-He2019b,CoB-He2021,CoB-He2022,CoB-Bu2023}
Co-Cu \cite{CoCu-Wang2017,CoCu-Wei2020,CoCu-Wei2023a,CoCu-Wei2023b}, and 
Co-Sn \cite{CoSn-Gaucherand2001,CoSn-Wang2015a,CoSn-Wang2015b,
CoSn-Qui2016,CoSn-Wang2017} melts.    
More specifically, working with supercooled melts, 
solidification of the product directly in the ferromagnetic phase was achieved 
despite the melting point being higher than the $T_{\mathrm{C}}$. 
Although the transition from a mobile liquid metal phase solidifying into a ferromagnetic 
state enhances the development of magnetic field effects, 
the non-equilibrium precipitation from a supercooled melt effects 
the crystallization process, thereby perturbing the crystal structure, microstructure, 
morphology, and properties of the products.
Herein, Co samples were directly crystallized in the ferromagnetic state under 
equilibrium conditions using a low melting Co-S eutectic.  

After describing the 
the methods and experimental arrangements, the macroscopic and microscopic structural 
analyses, along with magnetic and conductivity properties, are presented and discussed. 
Lastly, this work and its findings are summarized to indicate future extensions 
that are beyond the scope of this initial attempt.  For example, 
the well-defined facets observed in our products 
have not been reported elsewhere and provide an example of potential properties that 
may be uniquely accessed by an equilibrium precipitation in high magnetic fields.  

\section{Experimental Details}

\subsection{Furnace-Magnet}
A custom insert was designed and built so a \SI{1200}{\celsius} resistive furnace 
could be operated in an 89-mm, room-temperature-bore of an Oxford Instruments 
9.4~T NMR superconducting magnet, whose current was adjusted to provide a 
specific field value and then set in persistent mode for fixed field operation.  
The unit, whose details are described in the Appendix, consists of the integration 
of three main components, namely a water-cooled, heat-exchanging outer jacket, 
an interior resistive heater unit, and extra passive insulation and support. 

\subsection{Synthesis and Isolation of Co products}
Cobalt samples were synthesized using a Co-S binary 
flux \cite{Lin2012,Lin2013,Canfield-2020}. Stoichiometric amounts of 
Co (Cerac, 99.5\%) and S (Alfa Aesar, 99.9995\%) were precisely weighed to give 
a final composition of Co$_{61}$S$_{39}$. This ratio was selected to target 
the Co-rich side of the Co-S eutectic while fixing the liquidus temperature 
below \SI{1050}{\celsius} as a practical consideration, Fig.~\ref{fig:1}. Sample mixtures were placed 
in an alumina crucible with the low melting S on top, and the loaded reaction 
vessel was subsequently sealed in a quartz tube after 5 cycles of evacuation and 
flushing with Ar gas. The sealed tubes were subsequently loaded into a custom-designed 
furnace situated in the room-temperature bore of the superconducting magnet. 
Samples were heated at a rate of \SI{5}{\celsius}/min to \SI{1050}{\celsius}, 
held for 24 hours, cooled at a rate of \SI{1}{\celsius}/hr to \SI{900}{\celsius}, and 
then cooled to room temperature while remaining in the furnace. 
This process was performed while the field of the magnet was 
held constant at either $B = 0$, 3, or 9 T.  
It is important to note, the furnace-magnet system 
does not presently allow the sample to be 
rapidly removed for centrifuging to remove flux due to the 
large magnetic forces present in the field gradient region. 

In a fixed field, several synthesis runs were performed in the center of the homogeneous field 
region, where the field variation was less than 0.1\% over the synthesis region, and 
Co pieces extracted from the Co-S boule were labeled by the synthesis field. 
In one instance, the synthesis occurred $6.35\,\pm\,0.51$~cm 
below the center of the homogeneous field region, where the field and its gradient were 
approximately $8.8\,\pm\,0.1$~T and $0.10\,\pm\, 0.05$~T/cm. 
The resultant Co products are described in Table~\ref{tab:table1}.

\begin{center}
\begin{figure}[b!]
\includegraphics[width=3.375in]{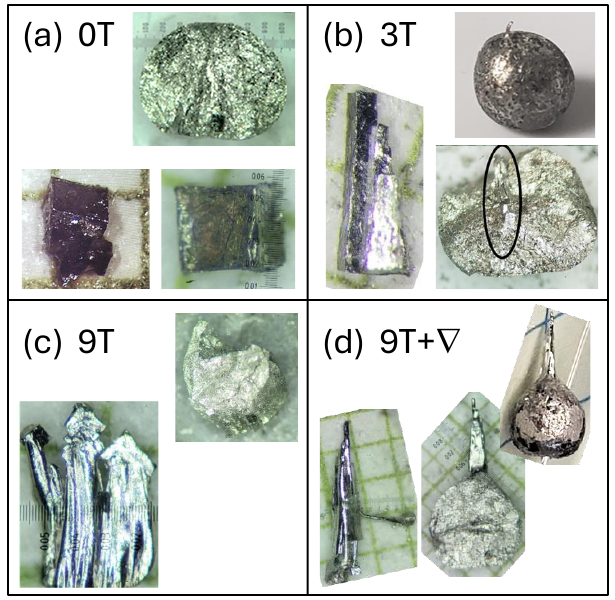}
\caption{Photographs samples described in Table~\ref{tab:table1}: 
(a) 0T - a cross-sectional view 
of about half the boule after first application of gentle force, where the characteristic 
flattened bottom is seen, and the small dark region is the Co crystal shown in the 
other two views after boule residue has been carefully removed; 
(b) 3T - the boule as grown, sitting on its bottom, with tip at the top noticeable, 
then a view looking down on the remaining boule after some processing to extract the Co piece, 
which is from the tip region circled in black, and shown as fully extracted; 
(c) 9T - a view of the boule, resting on its flattened bottom after some processing 
and the region, where the Co piece was extracted, is visible on the side near the 
bottom of the image, and the extracted Co piece is shown;
and (d) 9T+$\nabla$ - a side view of the boule as grown notably without a flatten bottom 
so it is laying on its side, followed by an image after 
some processing has been performed, and a view of the Co dendrite that was extracted for 
characterization, shown with a piece to the side due to the exceedingly soft nature of the 
resulting structure.  The background green paper grid, if present, provides a scale of 
1~mm, while in some micrographs an arbitrary scale bar from the lens is visible.}
\label{fig:2} 
\end{figure}
\end{center}

As shown in in Fig.~\ref{fig:2},reacted samples typically consisted of a boule of 
solidified flux surrounding 
the Co flux products. The latter were mechanically separated from the former through careful 
application of force to the relatively brittle flux matrix. Flux products were identified 
visually through their distinct, metallic appearance and clear facets in some cases. 
Pieces from the Co regions were extracted in sizes and orientations 
appropriate for the subsequent characterization technique.  

\begin{table}[b!]
\caption{\label{tab:table1}%
Product nomenclature, magnetic field and gradient during synthesis, 
saturation magnetization M$_{\mathrm{sat}}$, room-temperature resistivity, and 
RRR ($= \rho(300\,\mathrm{K}) / \rho(5\,\mathrm{K})$) values.}
\begin{ruledtabular}
\begin{tabular}{cccccr}
\textrm{Name} &
\textrm{$B$ (T)} &
\textrm{$\nabla B$ } & M$_{\mathrm{sat}}$  & $\rho(300\,\mathrm{K}$) & RRR\\
 &  & (T/cm) & ($\mu_{\mathrm{B}}/\mathrm{atom}$) & $\mu \Omega$~cm &  \\
\colrule
0T & 0 & $<10^{-3}$ & 1.66 & 11.0 & 2.4\\
3T & 3 & $<10^{-3}$ & 1.72 &  \;\,5.6 & 14.6\\
9T & 9 & $<10^{-3}$ & 1.40 &  \;\,5.6 & 2.6\\
9T+$\nabla$ & $8.8\,\pm\,0.1$ & $0.2\,\pm\,0.1$ & 1.61 & \;\,8.8 & 4.7\\
\end{tabular}
\end{ruledtabular}
\end{table}

\subsection{Characterization}
The X-Ray data were collected on a Bruker Dual micro source D8 Venture diffractometer 
and PHOTON III detector running APEX3 software package of programs and using 
Mo K$_{\alpha}$ radiation (0.71073~Å).
Phi scans with a sweeping angle of $0.1^{\circ}$  were performed on the 
sample at different $2\Theta$ angles to get Debye rings. The obtained Debye rings were 
integrated 
in APEX3 and converted into one-dimensional patterns.  Additional analysis was performed using  
 FullProf software \cite{FullProf} for Le Bail \cite{LeBail2005}
and Rietveld \cite{Rietveld1969} analysis.

Magnetic data were collected on a Quantum Designs MPMS XL SQUID Magnetometer with a magnetic 
range of $-7$~T to 7~T and a thermal range of 2~K to 300~K. The typical characterization 
sequence involved initial zero-field cooling (ZFC) to 5~K, measuring the 
temperature dependence of magnetization in 10~mT from 5~K to 300~K, and then 
measuring while field cooling (FC) to 5~K. 
Isothermal magnetization studies were performed by fixing the temperature and then sweeping 
the magnetic field from a low value to 7~T and then reducing the field while measuring to  $-1$~T. 
Samples were mounted in  No.~5 gelatin capsules and 
variable amounts of eicosane were used to preserve orientation. The magnetic contributions of 
both the capsule and eicosane were found to be less than the uncertainty of the total magnetic 
signal and were therefore considered negligible. Demagnetizing effects were also considered 
in respect to the geometry of the sample. Specific demagnetizing factors were calculated using the 
formula outlined by Bahl~\cite{Bahl2021} assuming a rectangular prism geometry. 
The overall shift due to demagnetization was minimal.

Resistivity data were collected with a Quantum Design PPMS using a 
Keithley 6221/2182A arrangement operating in delta mode over a 
range of 5~K to 300~K at a rate of 5 K/min. 
Four wires were attached to samples with silver paint in the Van der Pauw configuration 
for the 0T sample, and in a linear configuration for other samples as long, 
thin pieces were isolated from the products.

\section{Results and Discussion}

\subsection{Macroscopic Shapes/Properties}
The main piece of the zero-field control, referred to as 0T, appears 
to be a rectangular prism exhibiting multiple perpendicular, well-defined 
facets and sharp edges, Fig.~\ref{fig:2}(a). Notably, the sample 
morphology resembles a 
pseudo tetragonal habit with side-lengths within 10\% of each other 
and the third, shorter side ending in a jagged, poorly-defined face, 
suggesting an interrupted crystallization process in that direction. 
The overall appearance is consistent with the long-standing conception of 
Co adopting a fcc structure at high-temperature and suggests our heating 
profile successfully produced single crystals during their initial 
formation \cite{Honda1926,Kaya1928,Nishiyama1934,Myers1951,Sucksmith1954}, 
although the fcc to hcp transition is reported to split 
crystals \cite{Nishiyama1934,Sucksmith1954}. 
However, recently Sewak, Dey, and Toprek \cite{Sewak2022} have reported 
the hcp phase is stabilized at both low and high temperatures 
($\sim \SI{600}{\celsius}$) whereas the fcc phase is stabilized near \SI{227}{\celsius}, 
although the hcp phase was found to be the dominant low temperature phase 
in which fcc phase may be a sizable amount. 
Ultimately as will be discussed, all of the products 
reported herein possess a polycrystalline-like nature.

For the 3T sample, Fig.~\ref{fig:2}(b), clear facets and mutually perpendicular 
faces are evident in the flux products. However, the longest dimension has 
become distinct from the others by at least an order of magnitude, giving the product 
the appearance of an extended, rigid rod rather than a cube or box. 
This pattern is further evolved for the 9T sample, Fig.~\ref{fig:2}(c), where 
a mass of smooth, non-faceted columns of indeterminate morphology are capped 
with small, faceted rectangular prisms. 

Lastly, the morphology of the 9T+$\nabla$ sample, Fig.~\ref{fig:2}(d), possess 
additional different aspects as little to no faceting is present along 
a smoothly tapered dendrite that is more elongated relative to its thickness 
than the rods observed in the 3T sample, Fig.~\ref{fig:2}(a).  
Initially the dendrite was straight, like a needle, and embedded in a cluster 
much like the 9T product, Fig.~\ref{fig:2}(c). However, further processing 
revealed a cluster/bundle terminated in faceted feet. 
Carefully using tweezers, light pulling revealed the dendrite was 
exceptionally flexible but could be pulled out of the cluster, Fig.~\ref{fig:2}(d),  
although some plastic deformation resulted. 

\subsection{XRD Data and Interpretations}
\begin{figure}[t!]
\centering
\includegraphics[width=3.375in]{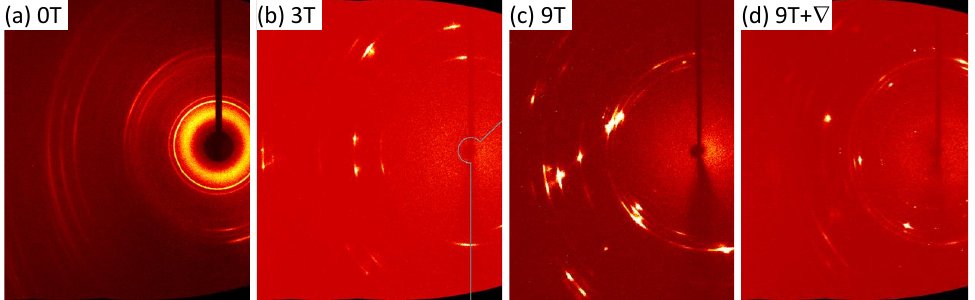}
\caption{Merged two-dimensional diffraction patterns 
collected on the samples listed in Table~\ref{tab:table1}.
The Debye rings possess polycrystalline-like features, 
with randomly oriented grains in the 0T sample.  
The samples grown in 
finite fields show clear evidence of significant texture.}
\label{fig:3}
\end{figure}

Merged two-dimensional diffraction patterns collected 
for the samples are shown in Fig.~\ref{fig:3}. 
The Debye rings show some evidence of 
polycrystallinity, with randomly oriented grains 
in the 0T sample.  
For synthesis in finite magnetic fields, 
the samples show a mix of rings and spots indicative 
of significant texture.

\begin{figure}[t!]
\centering
\includegraphics[width=2.80in]{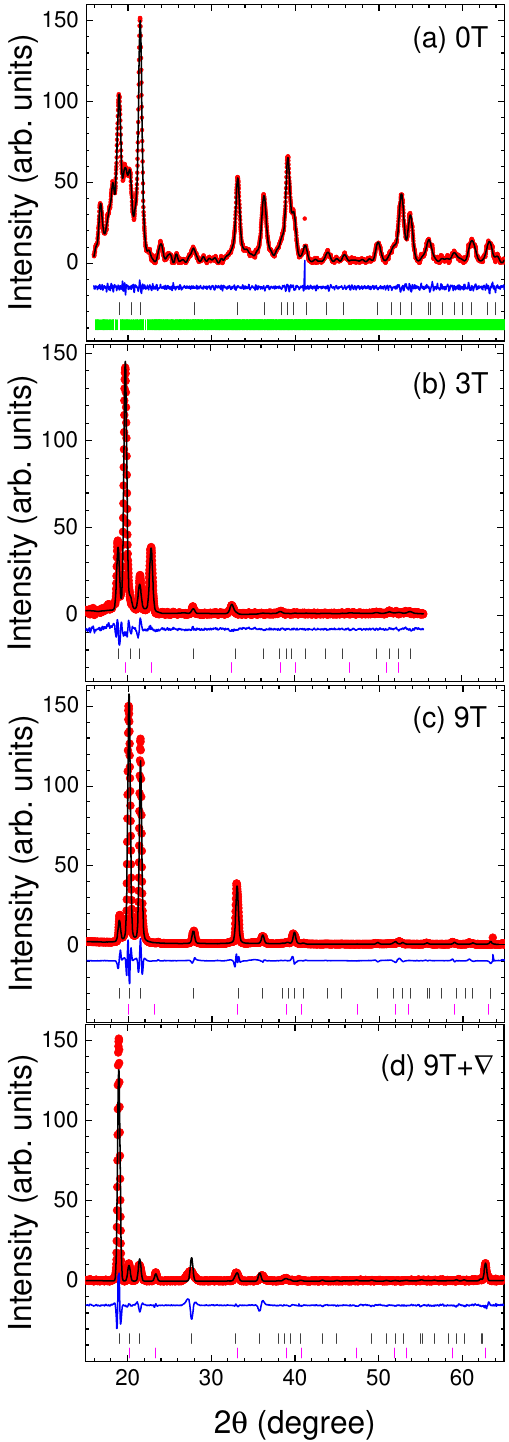}
\caption{The Le Bail fits of the integrated data from Fig.~\ref{fig:3}. 
Dots are the integrated data, and the black line is the best Le Bail fit. 
Below the data, the blue line is the difference, and the ticks represent:  
Co hcp (dary grey), Co fcc (magenta), and monoclinic S (green). The refined 
lattice parameters are reported in Table~\ref{tab:table2}.}
\label{fig:4}
\end{figure}

The results of the conversion of the data to total intensity versus 
$2\theta$ plots are shown in Fig.~\ref{fig:4}.
Several peaks with exceptionally large intensities and low peak widths 
(FWHM $< 0.05^{\circ}$) have been omitted for clarity and are 
attributed to artifacts of the merging and integration process, 
thereby resulting in some small breaks 
in the data curves.  The results of the Le Bail analysis are also shown, 
where a mixture of contributions from hcp and fcc phases of cobalt, 
along with monoclinic sulfur, provide suitable and plausible fits 
using the parameters given in Table II.

\begin{table}[b!]
\caption{\label{tab:table2}Lattice parameters used in the analysis of the XRD data, 
as contrasted to reference values.}
\begin{ruledtabular}
\begin{tabular}{ccccc}
 &\multicolumn{2}{c}{Hexagonal}&\multicolumn{1}{c}{Cubic}\\
 Name & a (\AA) & c (\AA) & a (\AA) & Ref.\\ \hline
 Co hcp & 2.5071 & 4.0686 &  & \cite{Taylor1950} \\
 Co fcc &        &        & 3.55 (3.54) &  \cite{Owen1954} (\cite{Taylor1950})\\
 0T & 2.504     & 4.031 &  $-$  & S \cite{Watanabe1974}\\
 3T & 2.510     & 4.036 & 3.601 & \\
 9T & 2.541     & 4.133 & 3.563 & \\
 9T+$\nabla$ & 2.486 & 4.057 & 3.530 & \\
\end{tabular}
\end{ruledtabular}
\end{table} 

Not surprisingly, the 0T control sample possesses Co hcp 
with a significant amount of monoclinic S, Fig.~\ref{fig:4}(a). 
For the synthesis in the presence of a magnetic field, the 
isolated products are generally comprised of a mix of 
hcp and fcc Co phases.  
Quantitative comparisons of these results with values reported 
in the literature are provided in Table~\ref{tab:table2}, 
where the outcomes from the Le Bail fitting are within 1-2\% 
of the published values for hcp Co and are reasonably close to 
the ones for fcc Co.  While the space group fits both 
structures, the obvious issue is the study was not 
performed on powder or single crystals, 
nevertheless the lattice parameters were resolved. 
In addition, 
the samples are not simply polycrystalline, so the estimate 
of the ratio between the two phases was not possible.

\begin{center}
\begin{figure}[tb!]
\includegraphics[width=3.375in]{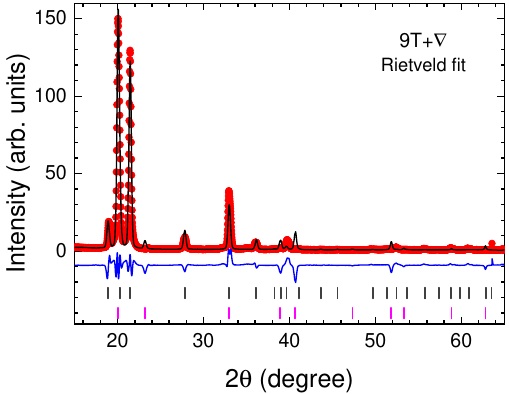}
\caption{The Reitveld fitting results for the 
9T+$\nabla$ data also used in Fig.~\ref{fig:4}(d).  
Dots are the integrated data, and the black line is the best Le Bail fit. 
Below the data, the blue line is the difference, and the ticks represent:  
Co hcp (dary grey) and Co fcc (magenta).}
\label{fig:5}
\end{figure}
\end{center}

In an attempt to further quantify the XRD data, Rietveld analysis 
was attempted for the three samples synthesized in a magnetic field, 
Fig.~\ref{fig:4}(b-d). Even with generous standards, the outcomes
for the 3T and 9T samples did not converge to 
physically plausible results. On the other hand, although 
not suitable for standard powder analysis, the 
Rietveld results for the 9T+$\nabla$ sample, 
Fig.~\ref{fig:5}, are reasonably the same as the outcome 
of the Le Bail fit, Fig.~\ref{fig:4}(d), with identical 
lattice parameters provided by both fitting 
schemes, Table~\ref{tab:table2}.

\subsection{Magnetic Properties}
\begin{center}
\begin{figure}[tb!]
\includegraphics[width=3.0in]{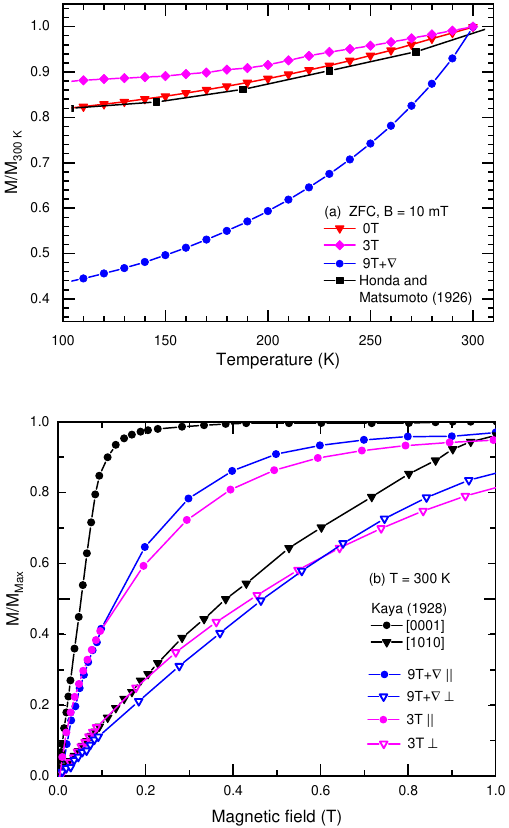}
\caption{The temperature and magnetic field dependences of the 
magnetic responses are shown. 
(a) The temperature dependences of the 
zero-field cooled (ZFC) responses of the magnetization, M, normalized to 
the value at 300~K, M$_{300\,\mathrm{K}}$, 
and measured while warming from 5~K, are shown over the temperature region 
that overlaps the data reported by Honda and Masumoto \cite{Honda1926}.
(b) The isothermal ($T=300$~K) magnetization, M, responses, normalized to the maximumum 
value observed/reported, M$_{\mathrm{max}}$, are shown over the field region 
that overlaps the data reported for a single-crystal of Co studied by 
Kaya \cite{Kaya1928}.  Parallel ($||$) and perpendicular ($\perp$) symbols 
refer to the direction of the measuring magnetic field compared to the 
growth direction along the field present during synthesis.}
\label{fig:6}
\end{figure}
\end{center}

The temperature and magnetic field dependences of the 
magnetic responses of the samples reported herein 
are shown in Fig.~\ref{fig:6}.  
The temperature dependence of the magnetization 
is similar for all samples except for the 9T+$\nabla$ response.
The isothermal magnetic field dependence of the magnetization 
suggests the 3T and 9T+$\nabla$ samples possess 
partial alignment of 
the magnetic easy-axis ([0001] or c-axis) of their hcp phases 
along the field direction 
present during the 
synthesis process.  

For the samples reported herein, 
saturation magnetization was achieved nominally about 3~T, 
independent of orientation of the sample, and the resulting 
magnetic moments are reported in 
Table~\ref{tab:table1}.
The $M_{\mathrm{sat}}$ values for all samples are in the range of 
$1.60 - 1.72~\mu_{\mathrm{B}}$ per atom, assuming the sample is pure cobalt.
These values can be contrasted with others reported in experimental 
studies \cite{Moon1964,Reck1969,Nishizawa1983,Chen1995} 
and theoretical/numerical works \cite{MPmain,MPmag,MPCoHCP,Tran2020}.

\subsection{Conductivity Measurements}
For an additional perspective on properties of the products, 
the resistivity as a function of temperature for 
several samples was measured, and in every instance, the data 
exhibited behavior typical for a simple metal, namely a near-linear temperature 
dependence from room temperature 
that smoothly transitions to a temperature-independent region  starting below nominally 30~K.
At room temperature, pure Co is reported to have a resistivity value of $5.2 - 5.9~ \mu \Omega$~cm \cite{Hall1968}, 
which was observed in the 9T and 3T samples, but 
larger values were obtained from the 
the 0T and 9T+$\nabla$ samples.
In addition, relatively low residual resistivity ratios,  
RRR $= \rho(300\,\mathrm{K}) / \rho(5\,\mathrm{K})$, were 
observed for all samples,  
but pure Co is notoriously known to possess rather low RRR values, 
typically reaching only $20 - 60$ \cite{Hall1968}.  
The resistivity results are summarized in 
Table~\ref{tab:table1}.
Bloch-Gruneisen fits~\cite{Nowotny2001} to the electrical resistivity data (not shown) give 
values of the Debye temperature of $480 \pm 30$~K, which is consistent with 460 K, 
which is the reported value for cobalt in the low temperature limit~\cite{Stewart1983}.

Ensemble, these data provide little definitive insight into the 
specific nature of the samples produced and investigated.  Nevertheless, 
benchmarks for future work are established, while also providing 
an important caveat to avoid cold working the samples after production.

\section{Summary}

Using an unusual combination of a furnace operating inside 
the room-temperature bore of a superconducting magnet, 
a striking suggestion to explore equilibrium materials 
synthesis in an exotic 
combination of parameter space was realized. 
Specifically, samples of Co were grown directly in the 
ferromagnetic state under equilibrium conditions using a 
cobalt sulfide flux, as suggested by Canfield and 
coworkers some years ago \cite{Lin2012,Lin2013,Canfield-2020}. 
Indeed, in equilibrium conditions while in the presence of 
magnetic fields up to 9 T, Co products were found to exhibit 
progressively elongated morphologies that were enhanced 
in the presence of a gradient field.  
The findings reported herein provide a preliminary step 
to designing explicit refinements required for the Co-S studies, 
while also facilitating the nucleation of other investigations 
targeting new and metastable phases of materials that are not accessible 
with present protocols and techniques.

\begin{acknowledgments}
The authors acknowledge enlightening conversations with Ryan 
Baumbach, Paul Canfield, Bill Malphurs, Theo Sigrist, and Kaya Wei. 

This work and the development of the high-field furnace insert 
and related instrumentation was supported by the U.S.~Department of 
Energy’s Office of Energy Efficiency and Renewable Energy (EERE) under 
the Industrial Efficiency \& Decarbonization Office (IEDO) 
award number DE-EE0009131.
The views expressed herein do not necessarily represent the views of the
U.S.~Department of Energy of the United States Government or any agency thereof. 

The 9.4~T magnet 
and the associated facility were supported by the 
National High Magnetic Field Laboratory (NHMFL or MagLab) funded by 
National Science Foundation 
(NSF) cooperative agreement DMR-1644779 and the State of Florida.  
Additional NSF support for the Research Experiences for 
Undergraduates (REU) was provided by DMR-1708410 (Spring 2022), 
DMR-1852138 (Summer 2022), MagLab DMR-1644779 (Fall 2022), 
and MagLab DMR-2128556 (Spring 2023, Fall 2023, and Spring 2024).

Major aspects of this work were interrupted by the pandemic, and the 
impact of perturbations of evolving personnel changes led to a delay 
in the generation of this manuscript.

\end{acknowledgments}

\appendix*
\section{Details of Furnace}

The cooling jacket is a double-walled cylinder constructed from two 316 stainless steel pipes 
selected as a non-magnetic material with favorable mechanical properties to resist damage from 
heat and pressure. The overall outer and inner diameters of the jacket are 76.2~mm (3~in.)~and 
63.5~mm (2.5~in.), respectively, and both cylinders are 1.59~mm (1/16~in.)~thick. 
This arrangement leaves 
an empty 3.18 mm cylindrical shell within the jacket through which closed-cycle, chilled water 
at $15 - \SI{20}{\celsius}$ is 
passed at $4\, \pm \, 0.15$~gpm. To ensure uniform cooling, a custom manifold splits 
the incoming water into eight 1.59~mm 
(1/16~in.)~I.D. turrets distributed uniformly 
around the bottom circumference of the jacket.  
The return chilled water exists a single 12.7 mm outlet at the top of the jacket. 
Pressure gauges, a flow meter, and Type-K thermocouples  monitor the condition of the chilled 
water at both the inlet manifold and the outlet port, thereby allowing for automatic 
shutdown if abnormal or dangerous conditions are detected. 

The furnace consists of a ceramic fiber heater (Watlow, VC400J12A) with an inner diameter 
of 12.7~mm (0.5~in.),~an outer diameter of 50.8~mm (2~in.), and a length of 304.8~mm (12~in.). 
The nickel chromium heating elements are arranged in two counter-wound solenoids embedded in 
ceramic fiber insulation. The furnace is positioned with its 
center 304.8~mm (12~in.) 
from the bottom of the cooling jacket, where the power leads 
exit. A hollow 
thermocouple port (I.D.~3.56 mm, 0.14~in.)~extends radially from the outer wall into the 
hollow core of the furnace halfway up its length. 
Protected with alumina insulation, a Type-N thermocouple 
traverses the length of the furnace from the bottom of the jacket to this port. 
The Type-N thermocouple was selected 
as a nonmagnetic alternative to Type-K with a similar operational 
temperature range (-270 – \SI{1260}{\celsius}) \cite{Tener2025}. 
The output is used by a programmable temperature controller 
(Watlow, F4SH-FAA0-01RG) to reach or maintain the desired heating rate and/or temperature. 
Current is supplied by 150~V, 3.5~A DC power supply (Kepco) to avoid potential damage 
associated with the magnetic forces that would be exerted on the heating element wires 
with alternating current.

The furnace is wrapped in a layer of 6.35~mm (0.25~in.)~thick ceramic fiber blanket which 
helps keep it centered, limits heat losses at the outside wall, and provides friction 
which prevents its position from shifting vertically within the cooling jacket. 
A 152.4~mm (6~in.)~cylinder of fiberboard insulation is also positioned underneath the 
furnace as additional support. The inner diameter houses an alumina tube with 
I.D.~15.9~mm (5/8~in.)~that protects the insulation from abrasion when samples are inserted 
or removed. This assembly rests on a cap fixed at the bottom of the cooling jacket 
leaving only the central bore open for access to the furnace chamber. Likewise, a 
column of ceramic fiber insulation with a central alumina tube sits on top of the 
furnace and fills the rest of the cooling jacket. During operation, thermal baffles 
made from thin, central alumina rods with fiberboard disks are inserted through 
the top and bottom alumina tubes to reduce heat loss.\\

\newpage

\bibliography{Flynn}

\end{document}